# A Search Strategy of Level-Based Flooding for the Internet of Things

**Tie Qiu, Yanhong Ding, Feng Xia * and Honglian Ma**

School of Software, Dalian University of Technology, Dalian 116620, China;
E-Mails: qiutie@dlut.edu.cn (T.Q.); dyh2010@gmail.com (Y.D.); mhl@dlut.edu.cn (H.M.)

* Author to whom correspondence should be addressed; E-Mail: f.xia@ieee.org;
  Tel.: +86-411-8757-1582; Fax: +86-411-8757-1567.

**Abstract:** This paper deals with the query problem in the Internet of Things (IoT). Flooding is an important query strategy. However, original flooding is prone to cause heavy network loads. To address this problem, we propose a variant of flooding, called Level-Based Flooding (LBF). With LBF, the whole network is divided into several levels according to the distances (*i.e.*, hops) between the sensor nodes and the sink node. The sink node knows the level information of each node. Query packets are broadcast in the network according to the levels of nodes. Upon receiving a query packet, sensor nodes decide how to process it according to the percentage of neighbors that have processed it. When the target node receives the query packet, it sends its data back to the sink node via random walk. We show by extensive simulations that the performance of LBF in terms of cost and latency is much better than that of original flooding, and LBF can be used in IoT of different scales.

## 1. Introduction

Internet of Things (IoT) can be defined as a new dynamic network of networks, where every daily object can communicate with each other. The IoT is a phenomenon resulting from the fact that an increasing number of physical objects now have the ability to connect to the Internet [1].

Radio-frequency identification (RFID) [2], wireless sensor network (WSN) [3] and real-time location systems (RTLS) [4] are important enabling technologies for the IoT [5]. Various types of sensor nodes are deployed in the IoT. Each sensor is an information source and different sensor nodes capture different information content. The sensor data is obtained in real-time. An essential and core technology for the IoT is the Internet. The information is transmitted in real-time through wired and wireless networks. Through RFID and RTLS, the IoT can control objects intelligently with cloud computing [6], pattern recognition [7] and so on. The IoT still has many architecture, technology and business challenges.

Sensor nodes in the IoT are often tiny resource-impoverished battery powered devices which have sensing, computation, and communication capabilities. As sensor nodes get energy from a battery source, their energy is limited and should be used judiciously. When people design protocols for the IoT, energy efficiency is an important consideration. Node discovery (also referred as information discovery or search processing) in the IoT is a widely studied problem in the context of networks [8–13]. Querying in sensor networks can be classified in the following ways: (a) based on the type of data–one-shot information or continuous streams; (b) based on the type of queries–pull-based (where the sink node issues the queries for information) or push-based (where the detection of events triggers notification to the sink node), or a hybrid of these two; and (c) based on the query process–structured (using an index or hash table) or unstructured (the search initiator has no clue about the target). Node discovery is a key problem especially in PULL and UNSTRUCTURED networks [14,15]. In PULL networks, the sink node searches for nodes that send emergency data, for example the temperature is too high. In UNSTRUCTURED network, the search initiator (the sink node) doesn't have any clue about locations of target nodes. Node discovery in networks is a one-shot query, and we mainly do research on the problem based on UNSTRUCTURED networks.

Three important search mechanisms for one-shot queries in UNSTRUCTRED networks are: flooding, controlled flooding (using expanding rings), and random walk. These searching mechanisms have been extensively studied in WSNs [8,16–20]. The purpose of node discovery is to find nodes that have emergency data. Because the situation is very urgent, the latency of searching mechanism must be very small. One problem of random walk is its high latency and this makes it unsuitable for delay sensitive applications. We mainly consider using flooding mechanism to search emergency nodes, but there are many problems with the basic flooding mechanism. One problem is that when the search initiator sends a query packet to the network, nodes which are not the target forward the query packet to all their neighbors when they receive it. The query packet is transmitted until its Time to Live (TTL) value becomes zero. There are too many duplicates of query packet in network. This makes for networks with heavy loads; also the nodes' energy is used up quickly. How to set a suitable TTL value is the key problem when sending query packets. Another problem is that basic flooding strategy can't find the shortest path from the sink node to a target node (least hops from the sink node to the target) without any control strategies. In order to find the shortest path, we should use some control strategies. The aim of this paper is to propose a searching strategy based on the basic flooding mechanism to find the shortest path from sink node to target.

In this paper, we thus propose a new search mechanism based on basic flooding, named Level Based Flooding (LBF), which not only reduces searching energy consumption compared to basic flooding, but also can find the shortest path between sink node and target node. A sensor node's *level* stands for the least hops to the sink node. For example, if the level of a sensor node is 5, it means that the sensor node can communicate with sink node in at least five hops. In LBF, when we deploy a network, the sink node broadcasts a level building packet to the whole network first. After this, all nodes in the network are divided into different levels according to the hops to the sink node, and the sink node collects the level information of each node and stores that in its memory. The level of each node is the minimum hops to the sink node. When the sink node wants to search for an emergency node, it broadcasts a search packet and sets the TTL value to the level of the target node. Sensor nodes receiving a query packet determine whether to rebroadcast the packet according to their neighbors' information - the percentage of neighbors that have processed the packet. The broadcast process combines the advantages of random walk and flooding. If the real-time percentage value is larger than a predefined threshold, the node only transmits the packet to one neighbor that hasn't processed the packet. When the target node receives the search packet, it sends its emergency data to the sink node via random walk within *level* hops. Data packets go to the sink node by the shortest way. Through this, search energy consumption is reduced a lot and the target can find the shortest path to the sink node. We assume that the network in our study case is static and there is only one sink node in the network.

The rest of the paper is organized as follows: Section 2 outlines some related work on searching in networks. Section 3 mainly describes the problem we solve in this paper and gives the network model. In Section 4, we present the principle of level flooding strategy and explain the strategy in detail. In Section 5, we design the algorithm for our LBF strategy. Section 6 provides the experimental results and compares our strategy with the basic flooding strategy. Finally, in Section 7 we conclude the paper.

## 2. Related Work

*2.1. Techniques Based on Random Walk*

In Simple Random Walk (SRW) [15], when a node receives a search packet, it selects one of its neighbors randomly and forwards the search packet to the selected neighbor. In Random Walk with k-Memory [RWM(*k*)] [15], when a node receives the search packet, it avoids recently visited *k* neighbors and randomly selects a neighbor from the remaining neighbors and forwards the search packet to it. If there are no unvisited neighbors, one of neighbors will be selected randomly. RWM(*k*) decreases the overhead compared to SRW by a constant factor, but does not change the basic behavior of SRW. Rachuri *et al*. [21] proposed three protocols viz., Several Short Random Walks (SSRW) search, Random Walk with Level Biased Jumps (RWLBJ) search, and Level Biased Random Walk (LBRW) search. The proposed protocols use a combination of random walk and Level Biased Walk (LBW) to search the target information. The basic idea of SSRW is to initiate several short random walks, one after the other until a target node is found. RWLBJ is essentially a random walk with periodic jumps based on LBW. In the random step phase, a fixed/variable number of steps are taken by the walk based on SRW. In the jump phase, a fixed/variable number of steps are taken by the walk based on LBW. The basic principle of LBRW is that, the search packet starts at the sink node and traverses in a random path to one of circumference nodes and traverses back to the sink node in a random path. All three random walks have better performance than SRW. Silva *et al*. [3] proposed

Non-Revisiting Random Walks (NRWs), which avoids re-visiting neighbors by selecting the next hop randomly among neighbors with the minimum number of visits. Non-revisiting random walks significantly improve upon simple random walks in terms of querying cost and load balancing, and the push-pull mechanism is more efficient than the push-only for query resolution. Hang *et al*. [14] proposed a biased walk that visits nodes near the search initiator before visiting more distant nodes. It resembles a spiral walk. The motivation is to find a nearest copy of the target information and this makes it more suitable for continuous type of queries than for one-shot queries. Avin *et al*. [17] proposed a biased walk in which unvisited nodes are given more priority than visited nodes and this improves the coverage of biased walk. The biased walk is mainly applicable to dynamic environments and requires that nodes know their visited and unvisited neighbors.

*2.2. Techniques Based on Flooding*

The basic flooding strategy is the original flooding method. In the basic flooding strategy, the search initiator only sends a query packet to the whole network without any control. Query packets are broadcast in the network quickly and there exist large numbers of duplications of the packet in the network. Packets only die when their TTL values become zero. This makes the whole network too heavily-loaded to do nothing, and nodes' energy is used up quickly, so we can't use the basic flooding strategy directly. Much work has been done to improve basic flooding.

In a wireless network, typical examples are DSR [22] and AODV [23]. In both protocols, nodes search for a target gradually in order to avoid flooding the entire network. The procedure in DSR is relatively simple. A node searches its one-hop neighbors first, and if the target is not found, the node then searches the entire network. AODV uses a different approach, whereby a node increases its radius linearly from an initial value until it reaches a predefined threshold. After that, a network-wide search has to be performed. Both of these expansion ring schemes assume that there are route caches residing in the nodes. In these two strategies, the searching efficiency is improved a lot only when the target is closer to the sink node, but when the target is far away from sink node, the efficiency is worse than in basic flooding.

There also has been some work on target discovery in WSNs. In Expanding Ring Search (ERS) [9], the search protocol avoids network-wide broadcast by searching for the target information with increasing order of TTL values. TTL limits the number of hops that can be searched from the search initiator. The main disadvantage of this protocol is that it resembles flooding in the worst case. Rachuri *et al*. [24] propose a protocol called Coverage based Expanding Ring Search (CERS) for energy efficient and scalable search in WSNs. The basic principle of CERS is to route the search packet along a set of ring based trajectories that minimize the number of messages transmitted to find the target information. In Gossip search [25], the source node broadcasts the search packet and all receivers either forward it with a *Gossip Probability p* or drop it with a probability *1-p*. The main disadvantage of Gossip search is that sending message to most of the nodes even when the target is located close to the source node. Han *et al*. [26] propose a dynamic probabilistic flooding strategy which uses the relationship between extinction probability and generation probability of the process to obtain a valid flooding probability based on the number of neighbor nodes in fixed probabilistic flooding. This strategy is more effective for various networks and has low requirement on nodes.

Acquired mechanism [8] combines random walk with controlled flooding for query resolution. In this approach, each intermediate node which receives the query packet, collects information from its *d*-hop neighbors and resolves some portion of query. However it is applicable for complex one-shot queries for replicated data. In these flooding strategies, they only consider how to avoid useless flooding. They don't consider the flooding process and don't propose strategies to reduce the rebroadcast times.

*2.3. Other Strategies for Special Purposes*

Lee *et al*. [27] proposed a distance-aware robot routing (DAR) algorithm, which focuses on how to pick the shortest path for the mobile robot by considering characteristics that are different from packet routing. This strategy uses flooding strategy to find the shortest geographic path. In the route discovery process, the node receiving the Route REQuest (RREQ) scans its Neighbor Distance Table (NDT) and updates the Cumulative Distance Information (CDI) with the distance between itself and the neighbor node sending the RREQ. By going through intermediate nodes to the destination, the CDI is updated with the cumulative distance values among the nodes. The nodes maintain the CDI cache to store the RREQ of the minimum CDI in order to send it to the destination. Chao *et al*. [28] proposed an alternative approach in which sensor nodes use the correlation between the node identifier and the index of model tuples assigned by the dispatching scheme to forward the unknown data over the paths in the network with the highest diversity. The simulations show that the proposed approach not only reduces the total power consumption of query resolution process, but also improves the classification success rate. Doss *et al*. [29] propose two methods for query resolution depending on the type of query. The two methods are the nearest perimeter approach that exploits perimeter aggregation for All-type queries, and the probabilistic right-hand sweep approach for ANY-type queries. Chakroaborty *et al*. [14] present a two level hierarchy for efficient service discovery of WSNs. Proximal neighborhood discovery is the prerequisite for service discovery followed by Optimal Service Discovery (OSD) which is based on the set of peers that a node should choose in order to utilize its requirements, instead of implementing all its required services itself. Andreou *et al*. [30] present a novel framework called MicroPulse+ for minimizing the energy consumption during data acquisition in WSNs. MicroPulse+ continuously optimizes the operation of T (Query Routing Tree) by eliminating inefficiency of data transmission and data reception using a collection of in-network algorithms. All above mentioned strategies use flooding as the basic step to solve their problems. They don't make many improvements on basic flooding. The complex data structure also reduces the search speed. Sensor nodes need to consume energy to maintain the information. Based on the above review of the state of the art, we mainly make improvements on basic flooding in the following aspects.

- In order to stop useless flooding after finding the target, a search strategy based on level is proposed, which divides the whole network into levels. The level of each node stands for the least hops to sink node. *TTL* of query packets is set to the level of target node.
- In the searching process, we combine random walk and basic flooding. In basic flooding, sensor nodes rebroadcast query packets automatically. In our strategy, sensor nodes decide how to process query packets based on real-time neighbors' information. They rebroadcast the query packet or transmit it to one neighbor according to rebroadcast threshold. We conduct simulations to get suitable threshold for networks with different scales.

- With our protocol, the energy consumption is reduced as much as possible. Data packets go back to sink node in the least hops. To avoid using one back path too much, data packets go back via random walk, which makes the network more balanced. Data transmission delay is restricted to match delay sensitive applications.

## 3. Problem Statement

### 3.1. Assumptions

This paper uses the following network model, which can be used for many IoT applications.

- All sensor nodes are uniformly deployed in a two-dimensional area. And there is only one sink node in network to collect data sensed by sensor nodes.
- Sensor nodes use Boolean sensing model. In this model, the sensing range of each node is a circle of radius *R* and all nodes in the sensing range can get its message, or nothing.
- Sensor nodes are nearly static, which means that there are no nodes added to network after finishing deployment, and sensor nodes are isomorphic. No sensor nodes leave the network except that their energy is used up. Sensor nodes are static after deployment.
- In the searching process, each sensor node only broadcasts each query packet once and there are few packet conflicts in the broadcasting process.

According to above assumptions, we have designed an application example, which is shown in Figure 1, which shows the plan of a department. The system is used to monitor the temperature of the department. Sensor nodes are deployed in the department and they are not moving any more. There is only one sink node in the application, which is used to collect sensor data and to transmit data to a PC.

**Figure 1.** An application example of IoT. The network is static and there is only one sink node in the network.

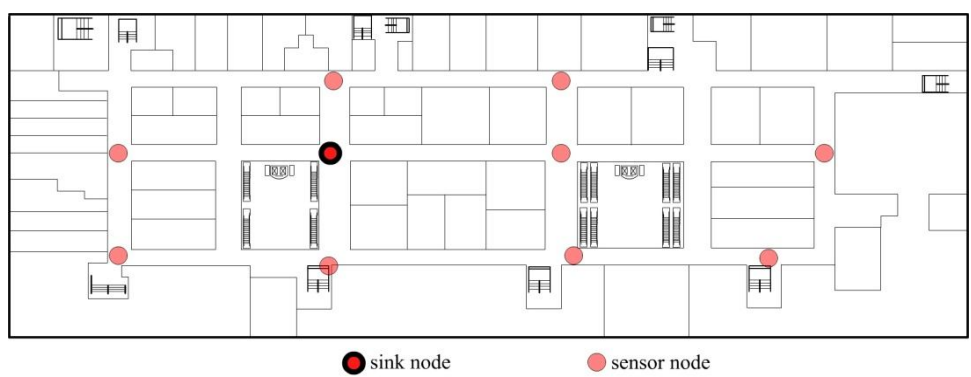

### 3.2. Problem Statement

The IoT can be used in environmental monitoring, industry control and so on. The data, such as temperature data sensed by sensor nodes, is called a resource. All data is transmitted to the sink node to be processed in the future. Once the sink node gets emergency data (for example, the temperature data is too high), it should search for the source node that sends the emergency data quickly and let the source node send more data to it quickly to determine whether there is emergency situation. The sink node sends a query packet to search for the target node. Sometimes the sink node may be interested in special events in the network. It may broadcast a query packet to the network to search for the node that detects an event. We should propose a strategy to transmit query packets to target nodes. When the target node gets a query packet, it should send back its data quickly to the sink node according to the search path.

As mentioned above, we consider using basic flooding strategy. As shown in Figure 2, when the sink node wants to search for target node, it sends a query packet. The query packet is diffused into the whole network and it diffuses in all directions from the sink node. As the communication radius of sink node is limited, only a few sensor nodes can receive the query packet through one hop. Each of those neighbors in turn rebroadcasts the packet exactly one time and this continues until all reachable network nodes have received the packet. We must reduce the redundant rebroadcast times as much as possible. When a target node receives the query packet, the query packet is still diffusing in the network until its TTL value is equal to zero. All sensor nodes should process the query packet at least once. In basic flooding, there are also loop transmission packets in the query process.

In the basic flooding strategy, if the target node wants to send data back, the query packet should record path information in the packet. As the query packet becomes larger and larger, it costs more and more energy to transmit the packet. The path information in the query packet is not the best path between the target node and the sink node. In Figure 2, the data packet may go back to the sink node via the yellow path, but the red path is a better solution.

**Figure 2.** Sink node searches for target node and the target sends data back to sink node. The query path is disordered and the data back path is not the best. Intermediate nodes rebroadcast the packet automatically.

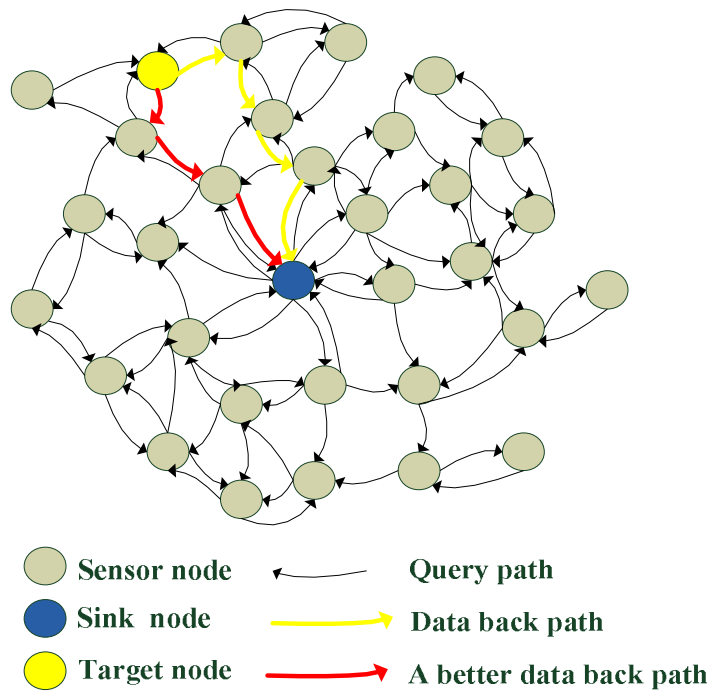

Therefore, when designing query strategy and data back strategy based on basic flooding strategy, we mainly consider the following three key points:

- The sink node doesn't have any location information about the target node;
- The strategy should reduce energy consumption as much as possible. We should also reduce redundancies as much as possible;
- The latency should be very low. As the situation is very urgent, the latency of finding target and the latency of transmitting data should also be very low.

In order to solve the above three problems, we propose a level based flooding strategy which is based on the basic flooding strategy.

*3.3. Analysis for Basic Flooding*

Basic flooding starts with a source node broadcasting a packet to all neighbors. Each of those neighbors in turn rebroadcasts the packet exactly one time and this continues until that all reachable network nodes have received the packet. There are two kinds of nodes in the network. One kind are Susceptible Nodes (SN), which haven't processed the message packet yet. Another kind is Infective Nodes (IN), which have processed the message packet. Only susceptible nodes respond to the message packet. When all susceptible nodes are around an infective node, the node is rendered dead and doesn't contribute further to the query packet spread. As the query packet diffuses in the network, there is a circle region of infected nodes centered at the source node (usually the sink node) which grows outwards with time, ultimately reaching the whole network. The spreading process is shown in Figure 3.

**Figure 3.** The message packet's spreading process in a network with basic flooding. The periphery of the infected nodes is shown by the full curve. The communication areas of nodes are shown by virtual curves.

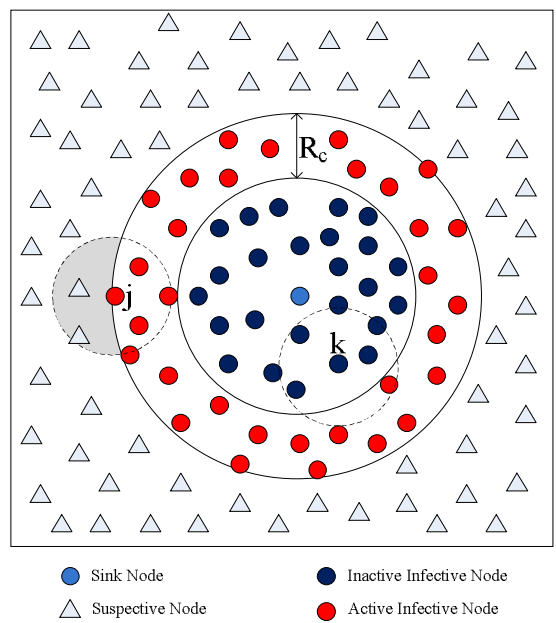

We approximate this observation by having nodes on the periphery of the infected circular region trying to infect their susceptible neighboring nodes lying outside the circle region. Only the infected nodes that lie within distance $R_c$ from the periphery of the circle of infected nodes (nodes with red color) can communicate with the susceptible nodes, and thus can push query packets forward. As shown in Figure 3, *Node k* can't infect any susceptive nodes, because all susceptive nodes lie outside its communication area. *Node j* can infective the susceptive nodes that lie in its communication area. Only the nodes in the shaded region can benefit from *Node j*'s rebroadcast, so a rebroadcast can only provide about 0~61% effective additional area [31]. The nodes in *Node j*'s rebroadcast ineffective area receive useless query packet, because they have processed the packet. We must reduce the redundant rebroadcast times as much as possible.

In the process of spreading query packets, each node receives packet duplicates from its all neighbors. The number of duplicates that every node receives depends on the node's degree. Figure 4 and Figure 5 prove this. As shown in Figure 4, *Node j* only has three neighbors, so it only receives three packet duplicates. Since *Node k* has eight neighbors, it may receive eight packet duplicates in each query process. *Node k*'s energy is consumed much faster than *Node j*'s.

**Figure 4.** Nodes with different degree consume energy at different speed. The speed is proportional to nodes' degree. *Node j*'s and *Node k*'s communication areas are presented by the virtual coils.

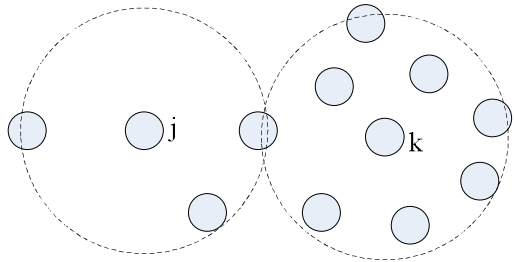

**Figure 5.** Each node's number of receiving packets is nearly equal to each node's degree. Each node's load in the basic flooding process is quite different. Nodes with high degree generate more traffic loads than nodes with low degree.

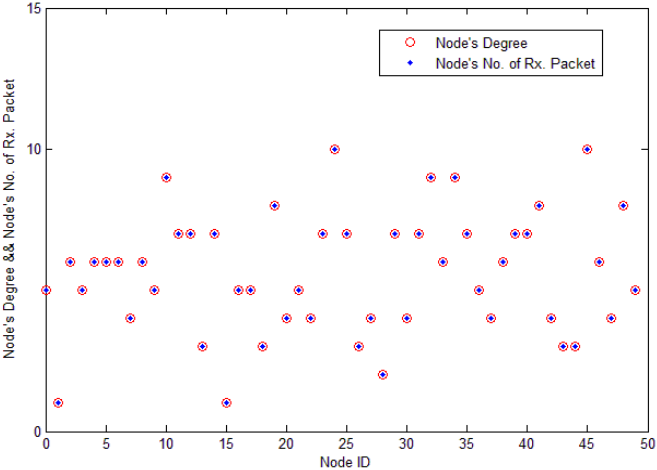

Load distribution in a network is usually very uneven due to factors such as network topology and network applications. In Figure 5, the x-axis stands for each node's ID, and the y-axis stands for each node's degree, which is the number of node's neighbors. If the value is 5, it means that the node has five neighbors and it receives five packets in each search process. From the figure, we see that each node's number of receiving packet is nearly equal to each node's degree. The speed of energy consumption is proportional to the nodes' degrees. As sensor nodes are randomly located in the monitoring area, each node's degree is quite different. Sensor nodes with large degree should reduce their rebroadcasts.

*3.4. Performance Metrics*

- *Average number of transmitted packets (i.e., cost of search) average_cost*: average number of packets transmitted for finding target nodes. When the sink node searches for target node $T_i$, the number of packets transmitted by all nodes is $C_i$. *average_cost* can be computed by Formula (1), where $N$ is the set of target nodes, and $|N|$ is the number of all target nodes:

$$average\_cost = \frac{\sum_{T_i \in N} C_i}{|N|} \tag{1}$$

- *Average energy consumed*: the average energy consumed for finding a target node. This includes the average number of packets transmitted and received for finding target nodes. When the sink node searches for target node $T_i$, the number of packet transmitted and received by all nodes is $EC_i$. *average_energy_cost* can be computed by Formula (2), where $N$ is the set of target nodes, and $|N|$ is the number of all target nodes:

$$average\_energy\_cost = \frac{\sum_{T_i \in N} EC_i}{|N|} \tag{2}$$

- *Average latency*: the average hops from the sink node to the target node. The number of hops between the sink node and one target node $T_i$ is $Hop_i$. *average_latency* can be computed by Formula (3), where $N$ is the set of target nodes, and $|N|$ is the number of all target nodes:

$$average\_latency = \frac{\sum_{T_i \in N} Hop_i}{|N|} \tag{3}$$

- *Rate of successful query*: the percentage that successful queries of total queries. The total number of queries is *TQnum*, and the number of successful queries is *Snum*. So the *suc_ratio* can be computed by Formula (4):

$$suc\_ratio = \frac{Snum}{TQnum} \times 100\% \tag{4}$$

- *Each node's average load for searching all targets*: each node's average number of packets transmitted and received for finding all target nodes. When the sink node searches for target node $T_i$, the number of packets transmitted and received by *Node q* is $Load_{i,q}$. So *average_load$_q$* can be computed by Formula (5), where $N$ is the set of target nodes, and $|N|$ is the number of all target nodes:

$$average\_load_q = \frac{\sum_{T_i \in N} Load_{i,q}}{|N|} \tag{5}$$

- *Convergence rate of level building process*: the time used for level building process. The time that the sink node sends the level building packet is $T_{start}$, and the time that all level building packets are dropped is $T_{end}$. *convergence_rate* can be computed by Formula (6):

$$convergence\_rate = T_{end} - T_{start} \tag{6}$$

- *The cost and energy consumption of level building process*: the cost means the total number of packets transmitted by all sensor nodes. The energy consumption includes the total number of packets sent and received by all sensor nodes. We assume that the cost and energy consumption of *Node $T_i$* in level building process is $LEC_i$, so the *EC_level_building* can be computed by Formula (7), where *N* is the set of target nodes:

$$EC\_level\_building = \sum_{T_i \in N} LEC_i \tag{7}$$

## 4. Modeling and Strategy

*4.1. Principle of Level Based Flooding*

In the basic flooding strategy, when the sink node wants to find a target node, it broadcasts a query packet to the whole network. Before finding the target, all nodes rebroadcast the packet automatically. This results in serious redundancy. One approach to alleviate above problem is to inhibit some nodes from rebroadcasting the packet to reduce the redundancy. When a target node receives the query packet, it responds to the sink node, but query packets are still flooded towards other directions continually. This is useless flooding. All sensor nodes process the query packet at least once. This wastes a lot of energy.

Sensor nodes may receive query packet duplicates from its neighbors that are closer to the source node before processing the packet. Nodes can determine whether to rebroadcast the packet according to the percentage of neighbors that have processed the packet. We define a threshold *P*, which stands for the percentage of neighbors that have processed the receiving packet, to limit the rebroadcast times.

The life time of query packet can be determined by its TTL value (the max hops that the packet can be transmitted). We can set the packet's TTL value properly to avoid unnecessary flooding. If TTL is too small, the query packet can't be flooded to target nodes; if TTL is too large, there will be useless flooding. The IoT is multi-hop network, and sensor nodes send their data to the sink node in multi-hops. We can set the query packet's TTL according to hops between target nodes and the sink node. We divide the whole network into different levels according to the hops between sensor nodes and the sink node. The nodes having the same hops to the sink node are in the same level. When the sink node wants to the search for the target node, it sets the packet's TTL to the level of target node.

*4.2. Brief Introduction of Level Based Flooding*

When we deploy the network, the sink node firstly sends a level building packet to the whole network. After the level building process is finished, the network is divided into concentric rings whose centers are the sink node. At the same time the sink node also gets the level values of nodes.

When nodes receive the level building packet, if the level information of node is updated, the node sends its latest level information back to the sink node. When the sink node wants to search for the emergency node, it broadcasts a query packet and sets the packet's TTL value according to the level of emergency node. When nodes who are not the target receive the query packet, if they have processed this packet, they drop it, or they decide how to process the packet according to the percentage of neighbors that have processed the packet. If the percentage of neighbors that have processed the packet is larger than predefined threshold, it only sends the packet to one of its neighbor that hasn't processed the packet. When the target gets the query packet, it sends its emergency data back to sink node in random walk within *level* hops. The back path may be different each time. The sink node can set TTL value appropriately according to the level information of target node, so the search packet doesn't always need to go through the whole network to find the target except that the target is in the edge of the network. This reduces the average energy consumption greatly. Because the network is divided into rings and level of each nodes is the minimize hops to the sink node, the emergency node can send its data to sink node within least hops. This reduces the latency and energy consumption. In the broadcasting process, sensor nodes don't always rebroadcast the packet, so the load of network is reduced a lot and the energy is saved.

Next we first introduce the overall design of LBF, and then we introduce each part in detail. In order to introduce the strategy conveniently, we give the following definitions.

**Definition 1.** The packet that sent by the sink node to divide the whole network into levels is called *LevelBuilding* packet.

**Definition 2.** The set that contains the node's neighbors whose level is lower than the node's level is called *LowLevelNeighbors* set.

**Definition 3.** The set that contains the node's neighbors whose level is equal to node's level is called *EqualLevelNeighbors* set.

**Definition 4.** The set that contains the node's neighbors whose level is higher than the node's level is called *HighLevelNeighbors* set.

*4.3. Overall Design of LBF*

The overall architecture of LBF is shown in Figure 6. From the figure we can see that LBF consists of four components: Level Building Process (LBP), sending Level Information Back (LIB), Searching Target Process (STP), Sending emergency Data Back (SDB). LBP is responsible for dividing the network into layers according to the distance between sensor nodes and the sink node. Level information of sensor nodes is sent to sink node by LIB. STP is used to diffuse the query packet to the whole network. At last the emergency data is sent to sink node by SDB. LBP and LIB run when we deploy the network and they run at the same time. STP and SDB run based on the results of the first two parts. Next we introduce each part in detail.

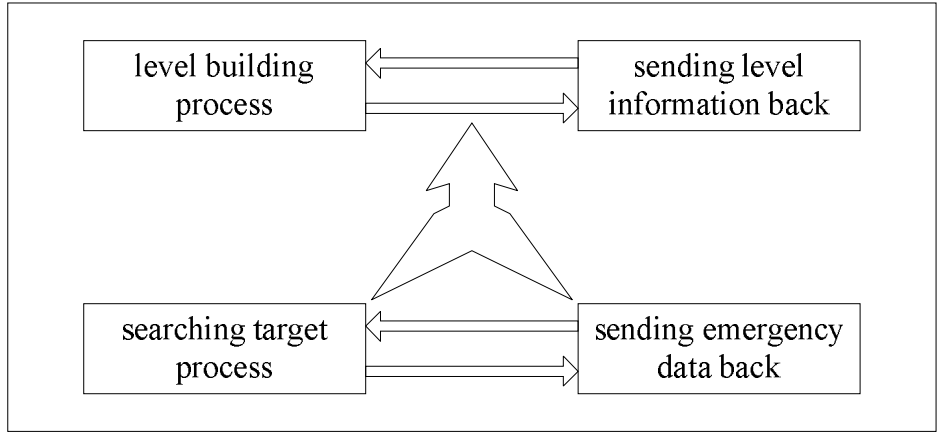

**Figure 6.** Architecture design of LBF.

*4.4. Level Building Process*

After deployment of the network, the sink node broadcasts a *LevelBuilding* packet by adding its level information to the packet. The level of the sink node is considered to be zero, so the initial level of *LevelBuilding* packet is zero. The level of sensor nodes is initialized to be a large integer, which is larger than the maximum level of the network. When a sensor node receives this packet, if *level* value in *LevelBuilding* packet is smaller than the node's *level*, the node updates its *level* as the level value in *LevelBuilding* packet plus one. Then the sensor node rebroadcasts the *LevelBuilding* packet by updating the packet's level with its level information. Sensor nodes send their level information to the sink node. How the information packet goes back to the sink node is introduced in Section 4.5. If the level value in *LevelBuilding* packet is larger than the node's *level* value or equal to the node's *level* value, the node drops the packet directly. A sensor node could receive more than one *LevelBuilding* packets, but it sets its *level* value to the minimum level of received *LevelBuilding* packets, so we can know that *level* value of each node stands for minimum hops between the node and the sink node.

In this process, each node also collects its neighbors' information, mainly including level information and node ID. When a node receives a *LevelBuilding* packet from its neighbor, if the neighbor's information is already in its cache and the information is the same as that in the cache, the node records nothing except deciding whether to rebroadcast the packet; if the node has nothing about the neighbor, it caches the neighbor's information including the node ID and its level information. After collecting all neighbors' information, nodes can segregate their neighbors into following three sets: *LowLevelNeighbors*: Set of all neighbors whose level values are smaller than that of the node; *HighLevelNeighbors*: Set of all neighbors whose level values are larger than that of the node and *EqualLevelNeighbors*: Set of all neighbors whose level values are equal to that of the node. *HighLevelNeighbors* set is used to broadcast the query packet. When a node broadcasts query packet, it only sends the packet to neighbors in *HighLevelNeighbors* set. *LowLevelNeighbors* set is used to find the shortest path to the sink node. When the node receives a back packet, it chooses a node from *LowLevelNeighbors* set as the next hop to make sure that the packet is closer to the sink node.

Level building process is only executed once and it is part of initial setup of the network. Since we assume a static network, there are no sensor nodes added to network and all nodes are static after deployment. When a node's energy is used up, it broadcasts a message to its neighbors to notice that it has died, and then its neighbors remove its information from the cache. The lifetime of *LevelBuilding* packet is not determined by its TTL. In the diffusion process, all sensor nodes rebroadcast query packet automatically. When all nodes' level value gets to the minimum value, this process finishes. The format of a level building packet is shown in Figure 7.

**Figure 7.** Format of level building packet.

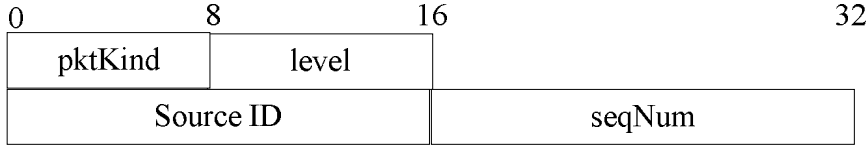

The meaning of each segment in Figure 7 is as follows:

- *pktKind:* the kind of packet;
- *level:* the hops that the packet has been transmitted;
- *Source ID:* the source node ID of this packet;
- *seqNum:* the ID of packet sent by the source. It is the unique identifier of packet;

Figure 8 shows the ring situation of level building process when the sink node is in center. It gives an example of a level building process. The level of the sink node is zero. The level of nodes in one hop to the sink node is level one and the level of nodes at two hops from the sink node is two. The level of nodes at three hops to the sink node is three, *etc*.

**Figure 8.** Result of level building when the sink node is in the center.

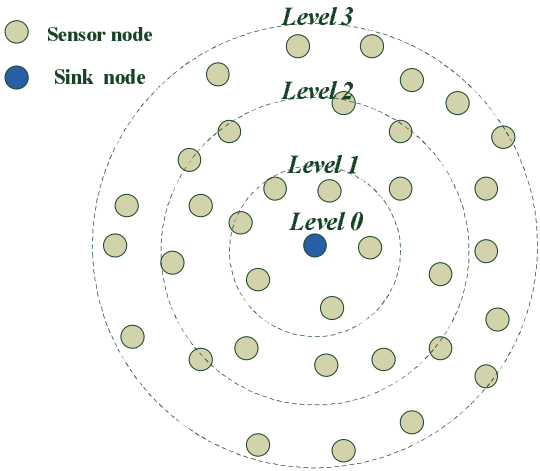

*4.5. Sending Back Level Information*

In LBF, when the sink node sends a query packet, it sets the packet's TTL value appropriately to reduce energy consumption as much as possible. The most ideal value is the least hops value between the sink node and the target node. The flood stops at finding the target. When a sensor node in the network receives a *LevelBuilding* packet, if the *level* in *LevelBuilding* packet is smaller than the node's *level*, the node updates its level value and it also sends its *level* back to the sink node at the same time. The node sends a reply packet which contains its level information. The reply packet's TTL is set to current level value of node. The reply packet goes back to the sink node in random walk within *level* hops.

The node has obtained its neighbors' level information in the level building process, especially neighbors who have lower levels. When a node receives a reply packet, if it isn't the sink node, it transmits the packet and chooses a node from *LowLevelNeighbors* set as the next hop of the reply packet. The reply packet gets closer to the sink node at each hop, and through *level* hops, the reply packet can get to the sink node. When the sink node receives a reply packet, if it doesn't have the node's level information, it records it; or it updates the node's level information. If the sink node receives several reply packets from a sensor node, level information in the latest reply packet is the best. The sink node records level information of all sensor nodes. The form of node information table is shown in Table 1.

**Table 1.** Level information table.

| Node ID | Level Information |
|---------|-------------------|
| 1       | 2                 |
| 2       | 5                 |
| 3       | 8                 |
| 4       | 1                 |
| 5       | 4                 |
| …       | ...               |

The format of a level back packet is shown in Figure 9.

**Figure 9.** Format of level back packet.

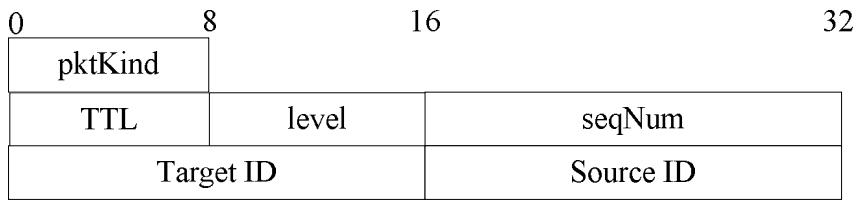

The meaning of each segment in Figure 9 is as follows:

- *pktKind:* the kind of the packet;
- *TTL:* the maximum hops of back packet that can be transmitted;
- *level:* the level value of source node;
- *Target ID:* the target node ID;
- *Source ID:* the source node ID of this packet;
- *seqNum:* the ID of packet sent by the source. It is the unique identifier of packet;

Figure 10 shows how sensor nodes send their level information back to the sink node. *Node n* selects a neighbor from its *LowLevelNeighbor* set as next hop of reply packet. All nodes receiving reply packet do the same thing as *Node n* does until the reply packet reaches to sink node. As shown in the figure, the back path is *Node n* → *Node m* → *Node k* → *sink node*. The level of *Node n* is 3; the level of *Node m* is 2; the level of *Node k* is 1.

**Figure 10.** Sensor node sends a reply packet to the sink node in a random walk.

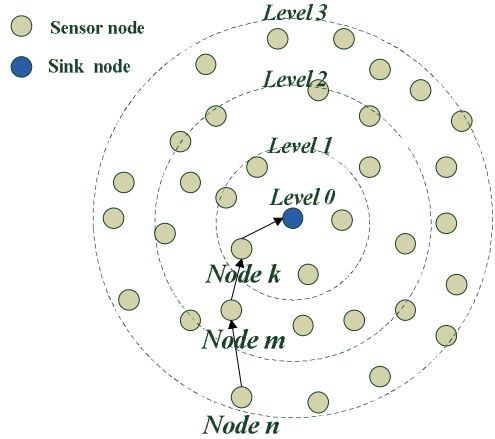

*4.6. Process of Searching Target*

If the sink node senses emergency sensor nodes, for example the sink node receives emergency data (a temperature is too high) from a node, the sink node should find the node quickly, or the sink node wants to find a node having special data. The sink node broadcasts a query packet to the whole network. The sink node sets TTL of the packet according to the target node's level information in its cache. Only sensor nodes in the zero level of TTL level need to process the query packet. Only when the node is far away from sink node, the query packet would be broadcast through the whole network to find it. If the node is very close to sink node, the searching cost and energy consumption will be reduced a lot. The average situation is much better than basic flooding. The node receiving a query packet decides how to process the packet according to the percentage of neighbors that have processed the packet to reduce rebroadcasts.

When receiving a previously unseen packet, an intermediate node can have three options to process it: rebroadcasting the packet, transmitting the packet to one of its neighbors that have not processed the packet or dropping the packet directly. Sensor nodes make localized decisions based on the packet's *TTL* value and parameter $p$, which stands for the real-time percentage of node's neighbors that have processed the receiving packet. The meaning of $p$ is shown in Figure 11. The query packet is diffused from the sink node. We can see *Node k* has seven neighbors. Some neighbors of *Node k* are closer to the sink node (gray nodes in Figure 11), so they may process the query packet before *Node k* in each query process. When *Node k* processes the packet, three of its neighbors have already processed the packet and four of its neighbors have not processed the packet, so $p$ can be computed by $3/7 = 0.428$.

The value of $p$ is computed in each packet diffusion process in real time. When the source starts a new query process, every node sets $p$ to zero. A node receiving a new packet initiates a counter $c$ that records the number of its neighbors that have processed the packet. Such a counter is maintained by each node for each query packet.

The node waits for a Random Assessment Delay (RAD), which is randomly chosen between 0 and $T_{max}$, before making decision on how to process the packet. Within this waiting time, the node may receive the same packet from its neighbors. When receiving a duplicate from neighbors during waiting time, the node updates $c$ plus one. When the waiting time expires, the node computes $p$ through $c/q$, where *q* is the total number of the node's neighbors. After getting *p*, the node decides how to process the packet.

**Figure 11.** Meaning of *p*. The spreading process is shown by full curve. The communication radius of *Node k* is presented by circle curve.

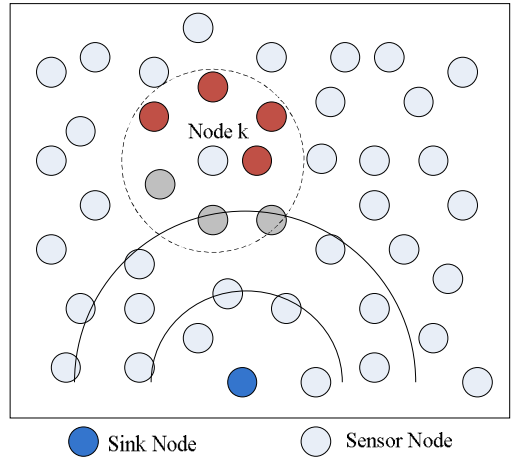

When a node receives a query packet, if its *hopCount* is equal to TTL, it drops the packet directly, or if it has processed the packet (every node records the packets that it has processed), it also drops the packet; otherwise it caches the query packet's basic information which is used to avoid reprocessing the query packet. The query packet is marked by sender's *ID* (the ID of node that starts the query) and the sequel number (*seqNum*) uniquely. If *hopCount* of the query packet is larger than level of the node, the node also drops the packet directly. Through above judges, each node only processes each query packet at most once.

If the node is the target, it sends its emergency data back to sink node quickly in random walk within *level* hops. If the node isn't the target, it makes decisions according to *p*. If *p* equals to one, which means that all neighbors have processed the packet, the node drops the packet directly. If *p* is larger than or equal to the predefined threshold *P*, the node transmits the packet to one of its neighbors that have not processed the packet. If *p* is smaller than *P*, it rebroadcasts the query packet to its neighbors with higher level, which means that only neighbors in *HighLevelNeighbors* set get the packet. The query packet won't be sent back to sink node in diffusion process. The *hopCount* value pluses one when the query packet is transmitted by a node. The packet dies when its *hopCount* value equals to TTL. And the query packet gets closer to target node with the increase of *hopCount*.

As shown in Figure 12, *Node j*, *Node k* and *Node l* have received the query packet and are going to make decisions to process the query packet. *Node l* finds that all neighbors have processed the packet, it drops the packet directly. *Node j* computes the percentage of neighbors that have processed the packet and the value is 0.67 which is larger than the threshold, so it only transmits the packet to a neighbor (*Node m*) that hasn't processed the packet. *Node k* computes the percentage of neighbors that have processed the packet and the value is 0.36 whose value is less than the threshold, so it rebroadcasts the packet.

**Figure 12.** Upon receiving the query packet, nodes make decisions on how to process the packet based on *p*. The periphery of diffusion area is shown by full curve. The communication areas of nodes are shown by virtual curves.

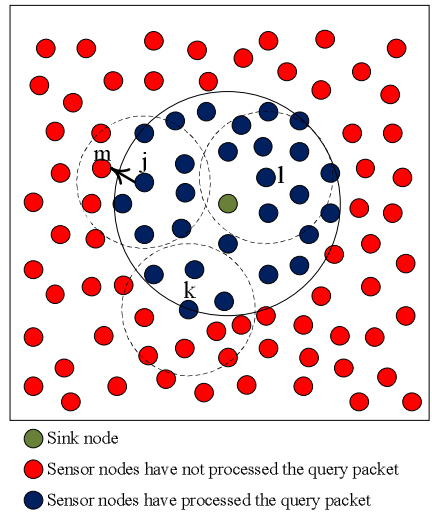

The format of a query packet is shown in Figure 13.

**Figure 13.** Format of query packet.

| 0 | 8 | 16 | 32 |
|---|---|---|---|
| pktKind | | | |
| hopCount | TTL | seqNum | |
| Target ID | | Source ID | |

The meaning of each segment in Figure 13 is as follows:

- *pktKind:* the kind of packet;
- *TTL:* the maximum hops of query packet;
- *Target ID:* the target node ID;
- *Source ID:* the source node ID of this query packet;
- *seqNum:* the ID of the query packet sent by the source. It is the unique identifier of query packet;
- *hopCount:* number of hops that query packet has been travelled.

Figure 14 shows how the query packet is transmitted to the target. The target node is *Node n*. The level of the *Node n* is 2, so TTL of the packet is 2. The node that starts the query is the sink node. When the sink node broadcasts a query packet, because the sink node's *level* is less than its all neighbors, it broadcasts the packet to its all neighbors. None of the nodes in level 1 is the target node, so they process the packet according to above mentioned making decision process. *Node n* is found in 2 hops. And the query packet stops at level 2. In this query process, only nodes in 2 levels process the packet. The query packet needn't to be broadcast in the whole network.

**Figure 14.** Query packet is broadcasted to *Node n*.

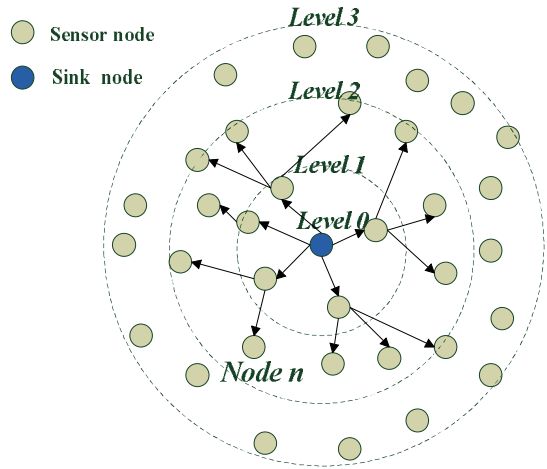

*4.7. Sending Back Emergency Data*

When the target receives a query packet, it sends its emergency data to the sink node quickly. This process is a little like the process introduced in Section 4.5. Emergency data packets get to sink node in a random walk within *level* hops. Every time the node that transmits packets chooses a node from its neighbors with a lower level as next hop of packets. The data packet gets closer to the sink node at each hop. When the hop num equals to TTL, the data packet also gets to the sink node. As packets get to sink node in a random walk, every packet may choose a different path so it can ensure the network's load is balanced and avoid using one path too much. If the back path is static, nodes in the path use up energy quickly. The lifetime of network should be reduced greatly. The format of a query packet is shown in Figure 15.

**Figure 15.** Format of data back packet.

| 0 | 8 | 16 | 32 |
|---|---|---|---|
| pktKind | | | |
| TTL | dataLen | seqNum | |
| Target ID | | Source ID | |
| Data | | | |

The meaning of each segment in Figure 15 is as follows:

- *pktKind:* the kind of packet;
- *TTL:* the maximum hops of packet;
- *datalen:* the total length of the data in the packet;
- *Target ID:* the target node ID;
- *Source ID:* the source node ID of this packet;
- *seqNum:* the ID of the packet sent by the source. It is the unique identifier of packet;
- *Data:* emergency data of the source node.

## 5. Design of the LBF Algorithm

According to Section 4, the LBF strategy consists of four parts: level building process, sending back level information, searching target process, and sending emergency data back. Next we design an algorithm for each part. The four parts are not independent: the first part and the second part run at the same time, and the third part and the fourth part run at the same time too.

The level building algorithm is shown in Algorithm 1. The sink node's level is initially set to be zero, and sensor nodes' level is initially set to be a large integer. The sink node broadcasts a level building packet. TTL of level building packet is infinite, which means level building packet won't die with the decrease of TTL. The process finishes when the level of nodes is unchangeable, which means that level of each node reaches the minimum value.

**Algorithm 1. LBP.**

> **Begin**
> *Step 1*. Node $N_i$ receives level building packet $P_i$ from Neighbor $Ne_i$.
> *Step 2*. Update $P_i.level$ plus one.
> *Step 3*. Determine whether to record $Ne_i$'s information.
>     *Step 3.1*. If $Ne_i$ is in neighbor list, go to Step 4.
>     *Step 3.2*. If $Ne_i.level < N_i.level$, go to Step 3.3; if $Ne_i.level == N_i.level$, go to Step 3.4; if $Ne_i.level > N_i.level$, go to Step 3.5.
>     *Step 3.3*. Store $Ne_i$ in *LowLevel_NeighborList*. Go to Step 4.
>     *Step 3.4*. Store $Ne_i$ in *EqualLevel_NeighborList*. Go to Step 4.
>     *Step 3.5*. Store $Ne_i$ in *HighLelve_NeighborList*. Go to Step 4.
> *Step 4*. Determine whether to update $N_i$'s level information.
>     *Step 4.1*. If $P_i.level < N_i.level$, go to Step 4.2; or go to Step 5.
>     *Step 4.2*. Update $N_i.level$, let $N_i.level = P_i.level$.
>     *Step 4.3*. Send $N_i.level$ back to sink node.
>             /*how the data goes back to sink node will be showed in Algorithm 2*/
>     *Step 4.4*. Rebroadcast level building packet $P_i$. Go to the End.
> *Step 5*. Drop the packet.
> **End**

LBP (Level Building Process) is used to divide nodes in network into different layers. Let each node's level reach to the minimize value, which means the least hops to the sink node. Next we give the algorithm that explains how the level information packet goes to the sink node, which is shown in Algorithm 2. The node sets *TTL* of reply packet to the node's level.

LIB (Level Information Back) shows how sink node get nodes' level information. Reply packet goes to the sink node in a random walk, and the random walk goes in levels. Reply packets can get closer to the sink node with the decrease of TTL. Through Algorithm 1 and Algorithm 2, the network has been divided into several layers and the sink node also knows the level information of each node. Next we propose the query algorithm, which is shown in Algorithm 3. When the sink node wants to search for emergency nodes, it broadcasts a query packet, and sets TTL of the packet according to the target node's level information.

**Algorithm 2. LIB.**

**Begin**
*Step 1.* Node $N_i$ receives reply packet $REP_j$ from Node $N_j$.
*Step 2.* Update $REP_j.TTL$ minus one.
*Step 3.* Determine whether to drop the packet.
  *Step 3.1.* If $REP_j.TTL$ != 0, go to Step 4.
  *Step 3.2.* Drop the packet, go to the End.
*Step 4.* Determine whether to transmit the packet $REP_j$.
  *Step 4.1.* If node $N_i$ is not sink node, go to Step 5.
  *Step 4.2.* If $N_j$ is in *NodeLevelInforList* and *level* < $REP_j.level$ , go to Step 4.3; or go to Step 4.4.
  *Step 4.3.* Update the level. Go to the End.
  *Step 4.4.* Store $N_j.level$ in *NodeLevelInforList*. Go to the End.
*Step 5.* Transmit the packet.
  *Step 5.1.* Choose a node from its *LowLevel_NeighborList* set as next hop.
  *Step 5.2.* Send the packet to the next hop node.
**End**

**Algorithm 3. STP.**

**Begin**
*Step 1.* Node $N_i$ receives the query packet $Q_i$.
*Step 2.* Update $Q_i.hopCount$ plus one.
*Step 3.* Initial the counter $c_x = 0$.
*Step 4.* Wait for RAD to expire.
  *Step 4.1.* Receive duplicate packet within RAD.
  *Step 4.2.* Update $c_x$ plus one.
  *Step 4.3.* Check whether RAD expires or not. If RAD expires, go to Step 5; else go to Step 4.
*Step 5.* Compute $p_x$.
*Step 6.* Determine whether to drop the packet $Q_i$.
  *Step 6.1.* If $Q_i.hopCount = Q_i.TTL$, go to Step 6.5.
  *Step 6.2.* If $Q_i.hopCount > N_i.level$, go to Step 6.5.
  *Step 6.3.* If $Q_i$ is in *ProcessedQueryPacketList*, go to Step 6.5.
  *Step 6.4.* Store $Q_i$'s information in *ProcessedQueryPacket*, go to Step 7.
  *Step 6.5.* Drop packet $Q_i$. Go to the End.
*Step 7.* Determine whether to transmit the query packet $Q_i$.
  *Step 7.1.* If Node $N_i$ is not the target node, go to Step 8.
  *Step 7.2.* Send emergency data back. Go the End.
      /*how to send back the packet will be showed in Algorithm 4*/
*Step 8.* Transmit the query packet.
  *Step 8.1.* If $p_x = 1$, drop the packet directly. Go to End.
  *Step 8.2.* If $p_x >= P$, transmit the packet to one of its neighbors that haven't processed the packet. Go to End.
  *Step 8.3.* If $p_x < P$, Send the packet to the neighbors in the *HighLevel_NeighborList*. /*avoid the query packet sent back*/
**End**

Searching Target Process (STP) shows how query packets are transmitted to the target node. According to the level information of target nodes, TTL of query packets can be set to the best value. The query packet also can be sent to a target node in the least number of hops. The node only broadcasts the packet to its neighbors in *HighLevelNeighbor* set, which can reduce the number of broadcast packets and let the packet get closer to the target node with the increase of *hopCount*. Next we give the algorithm that how the target node sends emergency data to a sink node. It is much same as Algorithm 2. It is shown in Algorithm 4. The node sets *TTL* of packet according to its level information.

Sending Data Back (SDB) shows how the target node sends its data back to sink node. Data packets go back to sink node in random walk within *level* hops. It may choose different path each time. This makes the load of network more balanced.

Much of the above algorithms' execution time is spent on searching for cached information (For example, when a sensor node receives a query packet, it should search cached information to determine whether it has processed the packet). If cached information is very little, each sensor node can process received packets quickly. Sensor nodes don't need to spend much time on processing packets, but only need to spend time on transmitting packet. This reduces the total time for transmitting packets to target node and reduces the energy consumption for processing packets.

**Algorithm 4. SDB.**

**Begin**
*Step 1*. Node $N_i$ receive reply packet $DREP_j$ from node $N_j$.
*Step 2*. Update $DREP_j.TTL$ minus one.
*Step 3*. Determine whether to drop the packet $DREP_j$
    *Step 3.1*. If $DREP_j.TTL$ != 0, go to Step 4.
    *Step 3.2*. Drop the packet, go to the End.
*Step 4*. Whether to transmit the packet.
    *Step 4.1*. If node $N_i$ is not the sink node, go to Step 5.
    *Step 4.2*. Get the data. Go the End.
*Step 5*. Transmit the packet.
    *Step 5.1*. Choose a node from its *LowLevel_NeighborList* as the next hop.
    *Step 5.2*. Send to the next hop node.
**End**

## 6. Experimental Results

We evaluate the performance of LBF and basic flooding protocols using NS-2, a popular discrete event network simulator. First we conduct simulations to get suitable thresholds ($P$) for networks of different scales. Then we compare the cost and latency of LBF against basic flooding.

*6.1. Simulation Setup*

In order to get comparative results of IoT of different scales, we have designed five simulation scenarios which are shown in Table 2. Sensor nodes are uniformly deployed in the monitored area and the sink node is nearly placed at the center of terrains. The communication radius of sensor node is 110 m [21,28]. We choose the value of 110 m according to most real applications.

Table 2. Simulation scenarios (communication radius = 110 m).

| Scenario No. | Number of Nodes | Square of Area ($m^2$) | Average Degree |
|---|---|---|---|
| 1 | 50 | 250 × 250 | 40.52 |
| 2 | 125 | 500 × 500 | 24.21 |
| 3 | 250 | 1000 × 1000 | 14.07 |
| 4 | 1000 | 2000 × 2000 | 9.26 |
| 5 | 4000 | 4000 × 4000 | 4.60 |

*6.2. Simulation Results of Level Building Process*

Table 3 shows the results of the level building process. With the increase of monitored area's square and the number of sensor nodes, the maximum level increases. The maximum level number means the maximum hops that are needed for a query packet to go through the whole network. The average hop number means the average hops for the sink node to find the target node at the best situation.

Table 3. Results of level building process.

| Scenario No. | Max Level | Average Level |
|---|---|---|
| 1 | 3 | 1.92 |
| 2 | 5 | 2.90 |
| 3 | 15 | 7.78 |
| 4 | 32 | 16.85 |
| 5 | 49 | 24.27 |

Figure 16 shows the convergence rate of level building process in each scenario. With the increase of network scale, the total time of the process becomes longer.

Figure 17 shows the cost and energy consumption of level building process. With the increase of network scale, the cost and energy consumption grows quickly. The number of processed packets is very large, because sensor nodes may process the *LevelBuilding* packet more than once.

**Figure 16.** Convergence rate of level building process in each scenario.

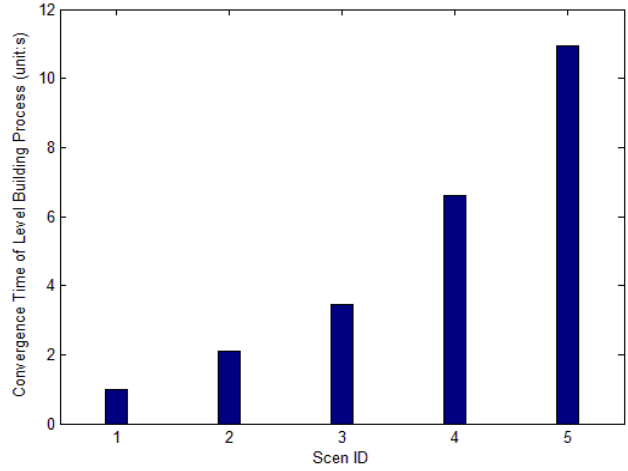

**Figure 17.** Cost and energy consumption of level building process.

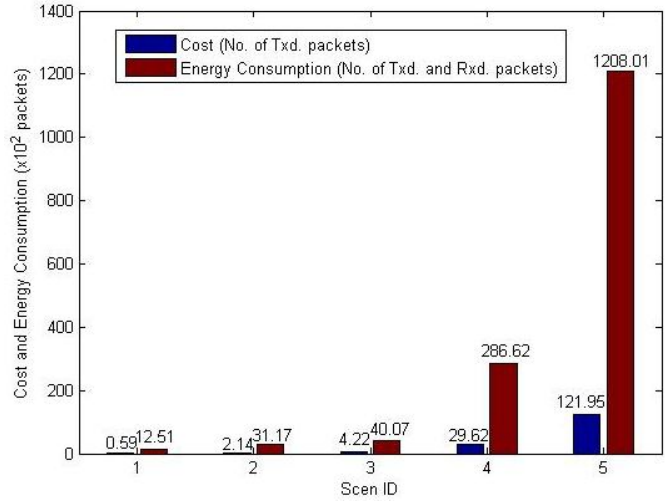

*6.3. Simulations to Get Suitable P*

In order to get suitable *P* for networks of different scales, we use the following three metrics:

- *Saved Rebroadcast (SR)*: *r* is the number of nodes receiving broadcast message, and *t* is the actual number of nodes that broadcast message packet. So *SR* can be computed by *(r−t)/r*.
- *Energy Consumption (EC)*: the total energy consumption in query process. The energy consumption is computed by the number of packets sent and received in the process. $e_i$ is the number of packets sent and received by *Node i* and *t* is the total number of nodes receiving the broadcast message. So *EC* can be computed by $\sum_{i=1}^{t} e_i$ .
- *REach ability (RE)*: the number of nodes receiving broadcast message divided by the total number of nodes that are reachable, directly or indirectly, from the source node. The total number of sensor nodes in network is *m*, and the number of nodes receiving broadcast message is *n*, so *RE* can be computed by *n/m*.

We conduct a series of simulations. The simulation results are shown in Figures 18–20. The performance of basic flooding can be found where $P = 1$. Each point in these figures represents our result obtained from a simulation run containing 1000 broadcast requests. Figure 18 shows the reach ability with different $P$ in every scenario. In Scenario 5, the reach ability grows with the increasing $P$. When $P$ is small, the reach ability is only about 10%. In other four scenarios, the reach ability in fact reaches about the same level as that of the basic flooding when $P$ is small. In high average degree network, $P$'s value doesn't affect the reach ability seriously, but in a low average degree network, $P$'s value has great effect on the reach ability because if the network's average degree is low, the nodes have few neighbors and it is too sparse to let the packet diffuse in the network.

**Figure 18.** Reach ability in each scenario against threshold $P$.

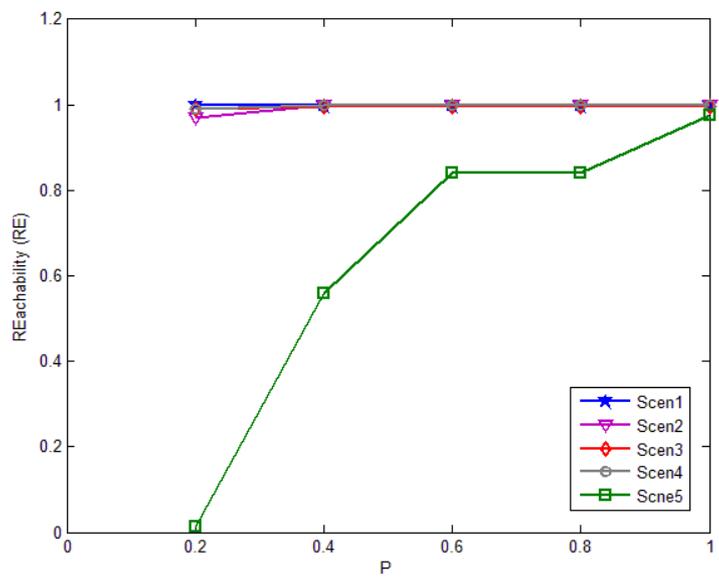

**Figure 19.** Saved rebroadcast in each scenario against threshold $P$.

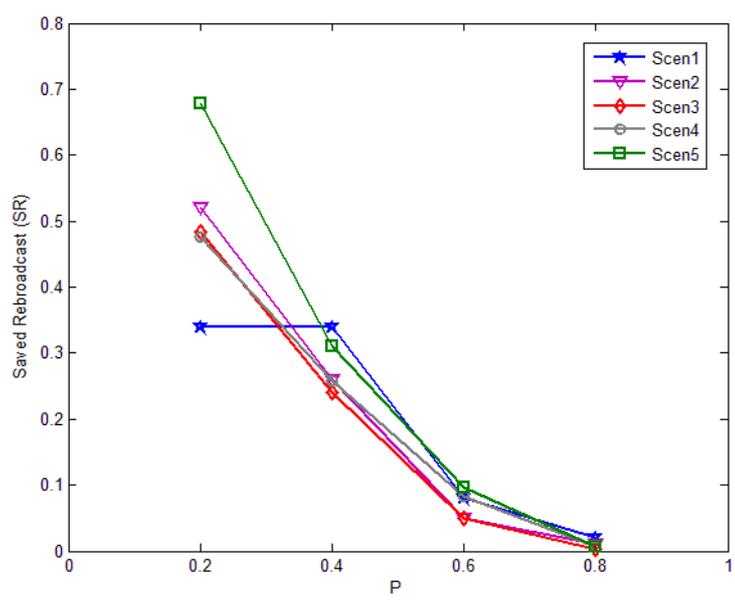

**Figure 20.** Energy consumption in each scenario against threshold *P*.

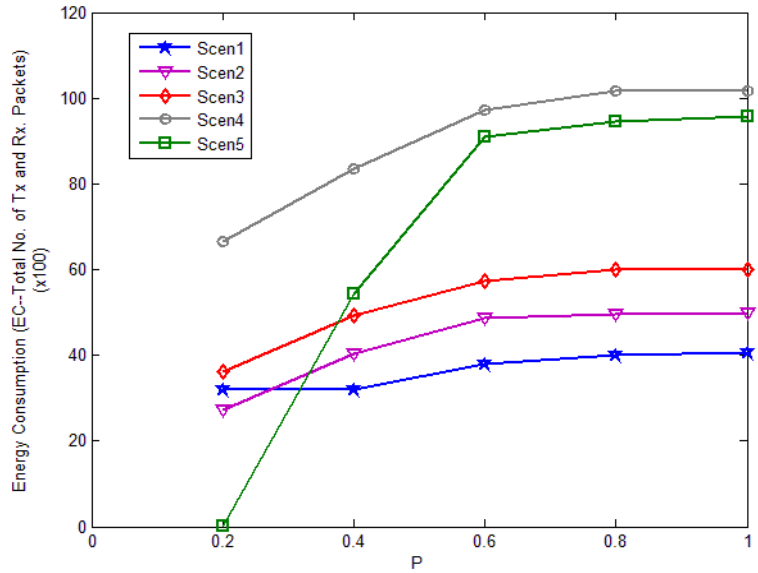

The saved rebroadcast in different scenarios with different *P* is shown in Figure 19. The amount of rebroadcast saving decreases proportionally to *P*, as *P* id increasing. When *P < 0.5*, the effect is obvious, and the scheme can save about 50%, but when *P > 0.5*, the scheme has nearly no effect. Various levels of saved rebroadcasts can be obtained over the scheme, depending on the average degree of nodes in the area. For instance, a scheme in lower degree (in Scenario 3) can offer about 50% savings at *P = 0.2*; while in a higher degree (in Scenario 1) it only offers about 30%.

Figure 20 shows the energy consumption in different scenarios with different *P* values. In every scenario, the energy consumption increases with the increasing *P*. When *P < 0.5*, the energy consumption increases obviously, especially in Scenario 5. When *P > 0.5*, the energy consumption is nearly the same as in the basic flooding case. The energy consumption in Scenario 5 is the best, but from Figure 18 we can see its reach ability is too small when *P = 0.2*, which affects the successful query rate a lot. From the above analysis, in the network with high average degree, threshold *P* can be set to 0.2 or 0.4, at which the network can get good reach ability and energy consumption. But in the network with low average degree, threshold *P* should be set larger than 0.5 to get high reach ability. Although the scheme with small *P* can offer good energy consumption, we can't ignore the scheme's reach ability. In the compare simulations, we set *P* to 0.4 in Scenario 1 and Scenario 2; we set *P* to 0.5 in Scenario 3 and we set *P* to 0.8 in Scenario 4 and Scenario 5.

*6.4. Performance of Searching Process*

Figures 21 and 22 are the simulation results of average load of each node to find all targets in LBF. The query packet is broadcast from nodes near the sink node to nodes far away from the sink node. If nodes are closer to the sink node, the average load is heavier, because they have to transmit more packets. From the two figures, we also can know the average load of each node is decreased with the increase of network scale.

**Figure 21.** Average load of each node in Scenario 2.

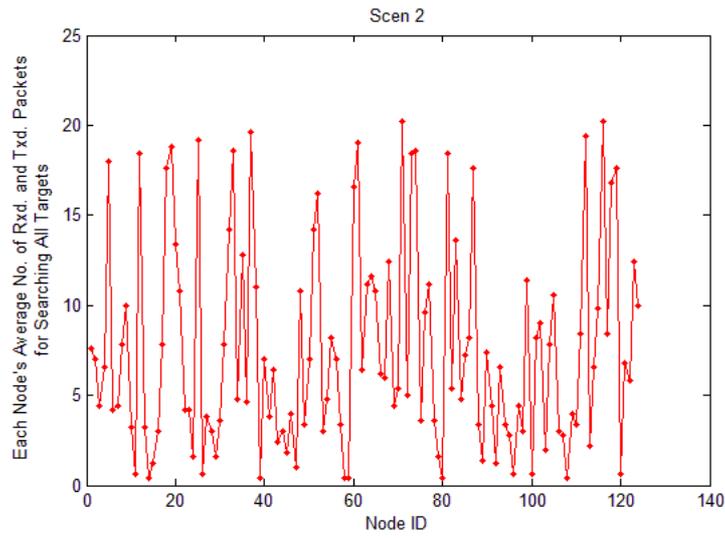

**Figure 22.** Average load of each node in Scenario 3.

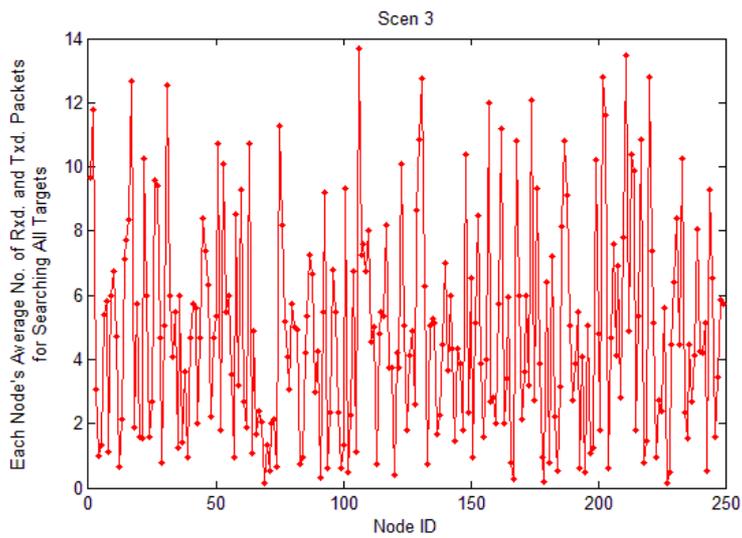

**Figure 23.** Fraction of nodes that have processed the query packet against the target node's level.

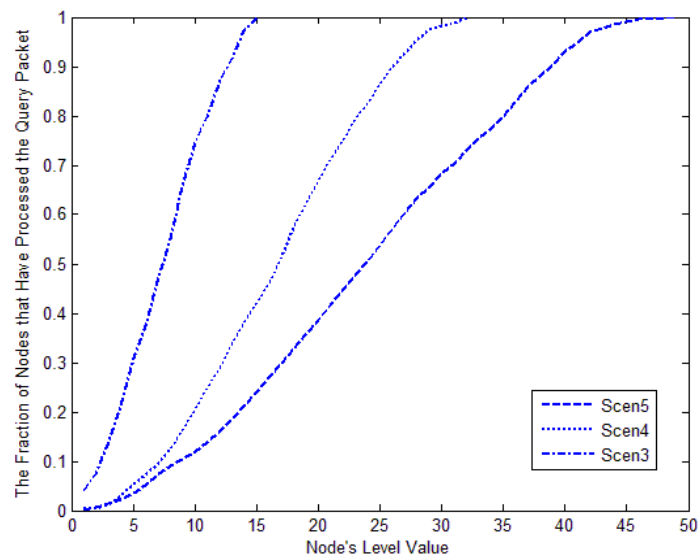

Figure 23 shows the relationship between the fraction of nodes that have processed a packet and the target node's level. From the figure we can see that the fraction of nodes that have processed a query packet increases with the target node's level. If the target node's level is higher, it should consume much more energy to search it. When the search packet's TTL is larger than nodes' maximum level, all nodes should process the packet. The situation is true in all three scenarios. The smaller the network is, the faster the curve goes up.

## 6.5. Performance Comparison

From Figures 24 and 25, we can know that the searching cost and energy consumption all increase with the improvement of network scale for each searching strategy. But the increase of LBF is much more slowly than that of basic flooding. Figure 24 shows the average cost of finding target nodes. We can observe that the average cost of search is much lower for LBF compared to that of basic flooding. The cost search is reduced about 54.5%. In basic flooding, every node nearly has to process a search packet at least once, which means a search packet needs to go through the whole network to find each target node. But in LBF, the average cost of search is nearly half. With the increase of nodes' number, the effect is much better.

**Figure 24.** Average cost of search to find target nodes.

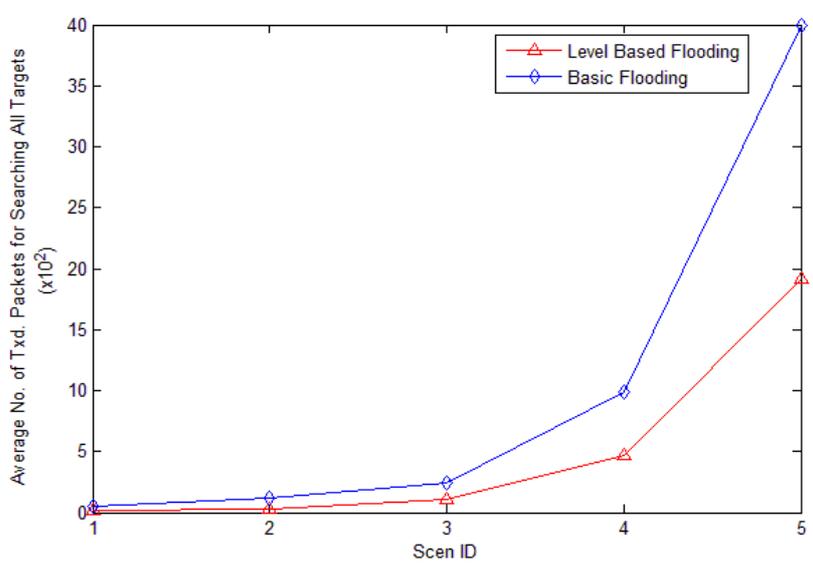

Figure 25 shows the average energy consumption for finding all targets. The energy consumption is about 54.4% of that of basic flooding. The average cost is nearly only half. Basic flooding strategy uses much more energy (more packets). With an increase of area square and number of nodes, the effect is more obvious. The trends of cures in Figures 24 and 25 are nearly the same. The searching cost and energy consumption are all much less than those of basic flooding.

**Figure 25.** Average energy cost for finding all targets.

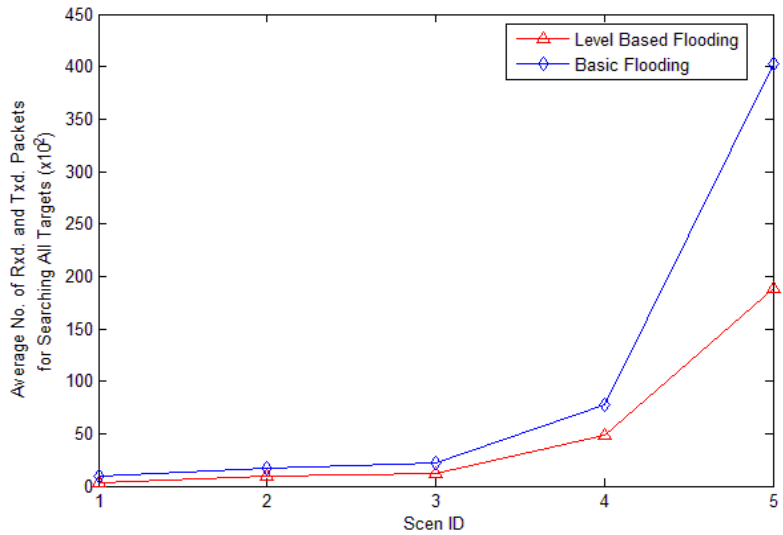

Figure 26 shows the average hops of search to find target nodes. We can observe that the average hops of search are nearly the same as the average level shown in Table 2. The average hops for LBF is less compared with that of basic flooding. If the number of data packets is larger, the energy consumption of transmitting data is much less than that of basic flooding. This can extend lifetime of network a lot. With the scale of network improved, the performance becomes better.

**Figure 26.** Average hops of search to find target nodes.

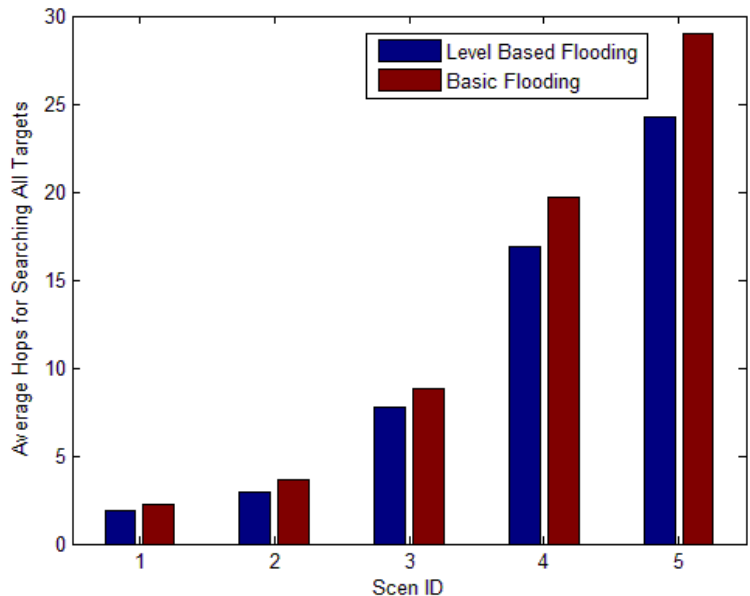

Figure 27 shows the ratio of successful query of the two strategies. The figure shows that the ratios of two strategies are all nearly 100%. As the nodes are all deployed randomly in monitored areas, there may be dead nodes exiting. As a consequence, the successful query ratio is not 100%. There is more control on the diffusion of query packets in our strategy, so the ratio is a little lower than that of basic flooding strategy.

**Figure 27.** Ratio of successful query.

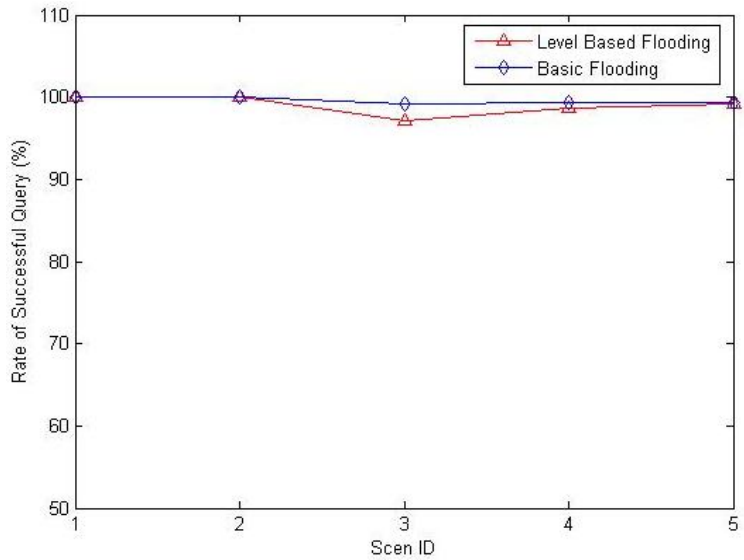

From Figure 28, we can see that no matter where the node is in network, all nodes should process the query packet in basic flooding. But with LBF, when the node is closed to the source node, only portion of nodes need to process the packet. When the node is in level 5, only 3.5% of nodes should process the packet; when the node is in level 40, 92.9% of nodes need to process the packet to search for the target.

**Figure 28.** Fraction of nodes that have processed the query packet in Scenario 5.

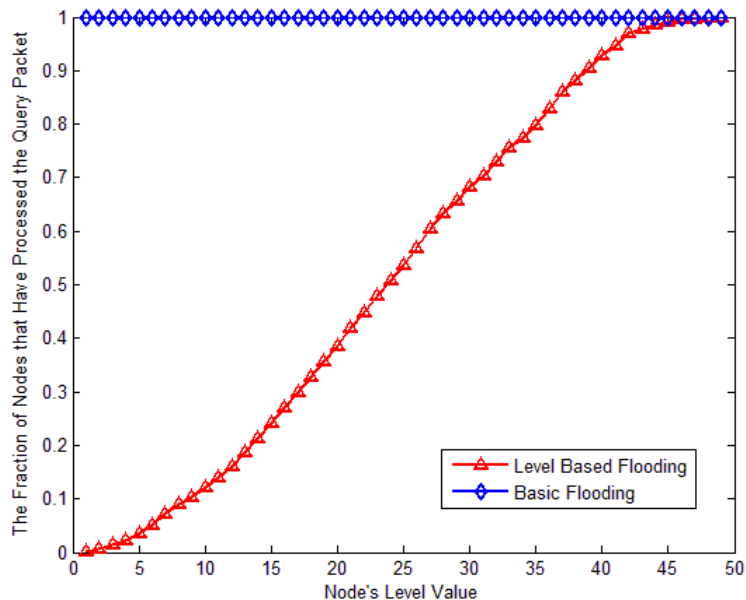

From the simulation results, we can summarize that our LBF search technique has less average searching cost, less average energy consumption and less average hops for finding a target node as compared against basic flooding. Most importantly, the above mentioned behavior is true in all five scenarios, therefore LBF can be used in IoT of different scales. Based on these results, we can clearly see that LBF is a better protocol as compared to basic flooding.

## 7. Conclusions

In this paper, we mainly focus on the searching problem of IoT. We have proposed LBF based on basic flooding, an energy efficient search or query resolution protocol. This protocol not only avoids useless flooding after finding the target, but also combines the advantages of random walk and flooding in searching processes. To the best of our knowledge, this method has never been exploited to address the searching problem in IoT. The basic principle of the protocol is to divide the whole network into different layers according to the hops between the sink node and sensor nodes. In the level building process, sensor nodes send their level information to the sink node and the sink node stores level information in its memory. When the sink node wants to find a target node, it broadcasts a query packet to the network and sets TTL of the packet according to the level information of the target. The packet is broadcast according to levels of nodes. Intermediate nodes receiving the packet decide whether to rebroadcast the query packet according to its neighbor information or to transmit the packet to one neighbor that has not processed the packet to reduce rebroadcasts. When the target node receives the packet, it sends the data back via random walk according to the levels of nodes. Data packets go back to sink node in the least hops and in variable paths. We show via extensive simulations that LBF can find the target node by incurring less cost and hops as compared to basic flooding. The results also show that LBF can be used in network systems of different scales. Our future work will be concentrated on the following directions:

- In this work, the network is static. There are no nodes added to the network after deployment. No nodes leave the network except that their power is used up. Sensor nodes don't move in the network. We intend to apply LBF in dynamic networks. We will develop a new strategy to process nodes' movement in the network. As nodes leave the network or are added to the network, we should take measures to process level changes of nodes.
- As query packets are broadcast from sensor nodes closer to the sink node, the nodes closer to the sink node consumes energy much faster than nodes at the edge of the network. The load of the network is unbalanced, which influences the lifetime of the network a lot. We intend to reduce the energy consumption of sensor nodes closer to sink node and make the load of network more balanced.
- The broadcast threshold $P$ in this paper is computed through simulations. That is, it is the empirical value. In future work, we intend to build a mathematical model for the threshold and get the best value for each network.
- In this paper, we only simulate our strategy on the NS-2 platform. In future work, we intend to build a read network to evaluate the performance of our scheme.

## Acknowledgments

This work is partially supported by Natural Science Foundation of China under Grant No. 60773213 and 60903153, Liaoning Provincial Natural Science Foundation of China under Grant No. 201202032, the Fundamental Research Funds for the Central Universities (DUT12JR10), the SRF for ROCS, SEM, and DUT Graduate School (JP201006).

**References**

1. Butgereit, L.; Coetzee, L.; Smith, A.C. Turn Me On! Using the "Internet of Things" to Turn Things On and Off. In *Proceedings of the 2011 6th International Conference on Pervasive Computing and Application*, Pretoria, South Africa, 26–28 October 2011; pp. 4–10.
2. Huang, Y.; Chen, Z.; Xi, J. A new RFID tag code transformation approach in internet of things. *J. Netw.* **2012**, *7*, 145–155.
3. Silva, J.S.; Krishnamachari, B.; Boavida, F. An Adaptive Strategy for Energy-Efficient Data Collection in Sparse Wireless Sensor Network. In *Proceedings of the European Workshop on Wireless Sensor Networks*, Coimbra, Portugal, 17–19 February 2010; pp. 322–337.
4. Ma, X.; Liu, T. The Application of Wi-Fi RTLS in Automatic Warehouse Management System. In *Proceedings of the 2011 IEEE International Conference on Automation and Logistics*, Dalian, China, 12–14 October 2011; pp. 64–69.
5. Mayordomo, I.; Spies, P.; Meier, F.; Otto, S.; Lempert, S.; Bernhard, J. Pflaum, A. Emerging Technologies and Challenges for the Internet of Things. In *Proceedings of the Midwest Symposium on Circuits and Systems*, Seoul, Korea, 7–10 August 2011.
6. Wang, H.; Yu, Y.; Zhu, P.; Yuan, Q. Cloud Computing Based on Internet of Things. In *Proceedings of the* 2011 *2nd International Conference on Mechanic Automation and Control Engineering*, Hohhot, Inner Mongolia, China, 15–17 July 2011; pp. 1106–1108.
7. Tu, B.; Yu, F. Bimodal emotion recognition based on speech signals and facial expression. *Adv. Intell. Soft Comput.* **2012**, *122*, 691–696.
8. Sadagopan, N.; Krishnamachari, B.; Helmy, A. The ACQUIRE Mechanism for Efficient Querying in Sensor Networks. In *Proceedings of the 1st IEEE International Workshop on Sensor Network Protocols and Applications*, Anchorage, AK, USA, 11 May 2003; pp. 149–155.
9. Chang, N.B.; Liu, M. Controlled flooding search in a large network. *IEEE/ACM Trans. Netw.* **2007**, *7*, 436–449.
10. Intanagonwiwat, C.; Govindan, R.; Estrin, D.; Heidemann, J.; Silva, F. Directed diffusion for wireless sensor networking. *IEEE/ACM Trans. Netw.* **2003**, *11*, 2–16.
11. Gehrke, J.; Madden, S. Query processing in sensor networks. *IEEE Pervas. Comput.* **2004**, *3*, 46–55.
12. Silva, J.S.; Krishnamachari, B.; Boavida, F. Querying Dynamic Wireless Sensor Networks with Non-Revisiting Random Walks. In *Proceedings of the European Workshop on Wireless Sensor Networks*, Coimbra, Portugal, 17–19 February 2010; pp. 49–64.
13. Gertz, M.; Ludascher, B. Optimizing Query Processing in Cache-Aware Wireless Sensor Networks. In *Proceedings of the 22nd International Conference on Scientific and Statistical Database Management*, Heidelberg, Germany, 30 June–2 July 2010; pp. 69–77.
14. Huang, H.; Hartman, J.; Hurst, T. Data-Centric Routing in Sensor Networks Using Biased Walk. In *Proceedings of the 3rd IEEE Communications Society Conference on Sensor, Mesh and Ad Hoc Communications and Networks*, Reston, VA, USA, 28 September 2006; pp. 1–9.
15. Ahn, J.; Kapadia, S.; Pattern, S.; Sridharan, A.; Zuniga, M.; Jun, J.H.; Avin, C.; Krishnamachari, B. Empirical evaluation of querying mechanisms for unstructured wireless sensor networks. *ACM SIGCOMM Comput. Commun. Rev.* **2008**, *38*, 17–26.

16. Cheng, Z.; Heinzelman, W. Flooding strategy for target discovery in wireless networks. *Wirel. Netw.* **2005**, 607–618, doi:10.1145/940991.940999.
17. Avin, C.; Brito, C. Efficient and Robust Query Processing in Dynamic Environments Using Random Walk Techniques. In *Proceedings of the 3rd International Symposium on Information Processing in Sensor Networks*, Berkeley, CA, USA, 26–27 April 2004; pp. 277–286.
18. Braginsky, D.; Estrin, D. Rumor Routing Algorithm for Sensor Networks. In *Proceedings of the 1st ACM International Workshop on Wireless Sensor Networks and Applications*, Atlanta, GA, USA, 28 September 2002; pp. 22–30.
19. Khan, M.; Gabor, A. An Effective Compiler Design for Efficient Query Processing in Wireless Sensor Networks. In *Proceedings of the Internal Conference on Circuit and Signal Processing*, Shanghai, China, 25 December 2010; pp. 157–159.
20. Chakroaborty, A.; Lahiri, K.; Mandal, S.; Patra, D.; Mrinal, K.; Mukherjee, A. Optimization of service discovery in wireless sensor networks. *Wired/Wirel. Int. Commun.* **2010**, *6074/2010*, 351–362.
21. Rachuri, K.K.; Murthy, C.R. Energy efficient and low latency biased walk techniques for search in wireless sensor networks. *J. Parallel Distrib.* **2011**, *71*, 512–522.
22. Johnson, D.B.; Maltz, D.A. *Mobile Computing, Chapter Dynamic Source Routing in Ad Hoc Wireless Networks*; Imielinski, K., Eds.; Kluwer Academic Publishers: Dordrecht, The Netherlands, 1996; pp. 153–181.
23. Perkins, C.; Royer, E.M. *Ad Hoc* on-Demand Distance Vector Routing. In *Proceedings of the Second IEEE Workshop on Mobile Computer Systems and Applications*, New Orleans, LA, USA, 25–26 February 1999; pp. 90–100.
24. Rachuri, K.K.; Muerthy, C.R. On the scalability of expanding ring search for dense wireless sensor networks. *J. Parallel Distrib.* **2010**, *70*, 917–929.
25. Haas, Z.J.; Halpem, J.Y.; Li, L. Gossip-based *ad hoc* routing. *IEEE/ACM Trans. Netw.* **2006**, *14*, 479–491.
26. Hou, N.; Feng, H.L.; Gu, X. Calculation of flooding probability based on number of neighbor nodes. *Appl. Res. Comput.* **2010**, *27*, 3443–3445.
27. Lee, I.H.; Yang, S.; Chao, S.H.; Kim, H.S. Robot path routing for shortest moving distance in wireless sensor network. *IEICE Trans. Commun.* **2010**, *E92-B*, 3495–3498.
28. Chao, C.H.; Li, I.H.; Yang, Y.; Li, J.S. An efficient diversity-driven selective forwarding approach for replicated data queries in wireless sensor networks. *J. Syst. Archit.* **2011**, *57*, 830–839.
29. Doss, R.; Li, G.; Mark, V.; Tissera, M. Information discovery in mission-critical wireless sensor networks. *Comput. Netw.* **2010**, *54*, 2383–2399.
30. Andreou, P.; Zeinalipour-Yazti, D.; Pamboris, A.; Chrysanthis, P.K.; Samaras, G. Optimized query routing trees for wireless sensor networks. *Inf. Syst.* **2011**, *36*, 267–291.
31. Mian, A.N.; Beraldi, R.; Baldoni, R. On the Coverage Process of Random Walk in Wireless Ad Hoc and Sensor Networks. In *Proceedings of the IEEE 7th International Conference on Mobile Ad hoc and Sensor Systems*, San Francisco, CA, USA, 8–12 November 2010; pp. 146–155.